\documentclass[aps,prb,twocolumn,showpacs,preprintnumbers,amsmath,amssymb,superscriptaddress]{revtex4}%

\usepackage{graphicx}%
\usepackage{dcolumn}
\usepackage{amsmath}
\usepackage{color}
\usepackage{multirow}

\begin{document}


\title{Fully gapped superconductivity in the topological superconductor $\beta$-PdBi$_2$}

\author{P. K. Biswas}
\email[Corresponding author: ]{pabitra.biswas@stfc.ac.uk}
\affiliation{Laboratory for Muon Spin Spectroscopy, Paul Scherrer Institute, CH-5232 Villigen PSI, Switzerland}
\affiliation{ISIS Pulsed Neutron and Muon Source, STFC Rutherford Appleton Laboratory, Harwell Campus, Didcot, Oxfordshire, OX11 0QX, UK}
\author{D. G. Mazzone}
\affiliation{Laboratory for Neutron Scattering and Imaging, Paul Scherrer Institut, CH-5232 Villigen, Switzerland}
\author{R. Sibille}
\affiliation{Laboratory for Scientific Developments and Novel Materials, Paul Scherrer Institut, CH-5232 Villigen PSI, Switzerland}
\author{E. Pomjakushina}
\affiliation{Laboratory for Scientific Developments and Novel Materials, Paul Scherrer Institut, CH-5232 Villigen PSI, Switzerland}
\author{K. Conder}
\affiliation{Laboratory for Scientific Developments and Novel Materials, Paul Scherrer Institut, CH-5232 Villigen PSI, Switzerland}
\author{H.~Luetkens}
\affiliation{Laboratory for Muon Spin Spectroscopy, Paul Scherrer Institute, CH-5232 Villigen PSI, Switzerland}
\author{C.~Baines}
\affiliation{Laboratory for Muon Spin Spectroscopy, Paul Scherrer Institute, CH-5232 Villigen PSI, Switzerland}
\author{J. L. Gavilano}
\affiliation{Laboratory for Neutron Scattering and Imaging, Paul Scherrer Institut, CH-5232 Villigen, Switzerland}
\author{M. Kenzelmann}
\affiliation{Laboratory for Scientific Developments and Novel Materials, Paul Scherrer Institut, CH-5232 Villigen PSI, Switzerland}
\author{A.~Amato}
\affiliation{Laboratory for Muon Spin Spectroscopy, Paul Scherrer Institute, CH-5232 Villigen PSI, Switzerland}
\author{E.~Morenzoni}
\affiliation{Laboratory for Muon Spin Spectroscopy, Paul Scherrer Institute, CH-5232 Villigen PSI, Switzerland}
\date{\today}

\begin{abstract}
The recent discovery of the topologically protected surface states in the $\beta$-phase of PdBi$_2$ has reignited the research interest in this class of superconductors. Here, we show results of our muon spin relaxation and rotation ($\mu$SR) measurements carried out to investigate the superconducting and magnetic properties and the topological effect in the superconducting ground state of $\beta$-PdBi$_2$. Zero-field $\mu$SR data reveal that no sizeable spontaneous magnetization arises with the onset of superconductivity implying that the time reversal symmetry is preserved in the  superconducting state of $\beta$-PdBi$_2$. Further, a strong diamagnetic shift of the applied field has been observed in the transverse-field (TF) $\mu$SR experiments, indicating that any triplet-pairing channel, if present, does not dominate the superconducting condensate. Using TF-$\mu$SR, we estimate that the magnetic penetration depth $\lambda=263(10)$ nm at zero temperature. The nature of $\lambda(T)$ provides evidence for the existence of a nodeless single \textit{s}-wave type isotropic energy gap of 0.78(1) meV at zero temperature. Our results further suggest that the topologically protected surface states have no effect on the bulk of the superconductor.
\end{abstract}

\maketitle


A topological superconductor is characterized by a full pairing gap in the bulk and topologically protected gapless surface states that are essentially Andreev bound states consisting of Majorana fermions. The recent discovery of topologically protected surface states in superconducting $\beta$-PdBi$_2$ has generated great interest in this material as a perfect platform to study the topological aspect in the superconducting pairing state~\cite{Sakano}. PdBi$_2$ crystallizes into two different phases depending on different growth conditions. The $\alpha$-phase stabilizes below 380$^o$C and the $\beta$-phase is realized between 380$^o$C and 490$^o$C~\cite{Matthias, Okamoto}. Both $\alpha$-PdBi$_2$ and $\beta$-PdBi$_2$ are superconducting below 1.73 and 4.25 K, respectively~\cite{Zhuravlev1}. Few other members of this Pd-Bi family also show superconductivity, these are e.g. $\alpha$-PdBi with a $T_{\rm c}$ of 3.8 K~\cite{Zhuravlev1} and $\gamma$-Pd$_{2.5}$Bi$_{1.5}$ with a $T_{\rm c}$ of 3.7-4 K~\cite{Zhuravlev2}. Recent studies have shown that the effective superconducting transition temperature of $\beta$-PdBi$_2$ can be enhanced from 4.25 to 5.4 K by improving the sample quality~\cite{Imai}. A positive curvature in the temperature dependent upper critical magnetic field and a second hump in the specific heat data suggest that $\beta$-PdBi$_2$ is a multiple-band/multiple-gap superconductor~\cite{Imai}. However, the specific heat data were limited to a minimal temperature of 2 K. Scanning tunneling microscopy (STM) results suggest that $\beta$-PdBi$_2$ is a single-gap multiband superconductor~\cite{Herrera}. Theoretical calculations of the electronic band structure and the Fermi surface (FS) further suggest different energy gaps on distinct FS sheets~\cite{Shein}. Detection of topologically protected surface states in $\beta$-PdBi$_2$ using (spin-)angular-resolved photoemission spectroscopy, (S)ARPES has placed this material in the class of topological superconductors~\cite{Sakano}. The interesting aspect of this new finding is that topological superconductivity may be realized in a pure stoichiometric compound. In other proposed candidates such a state can be achieved exclusively by a carrier doping, e.g. Cu-intercalated Bi$_2$Se$_3$~\cite{Hor,Kriener,Fu,Sasaki1} and In-doped SnTe~\cite{Sasaki2} or by applying pressure, e.g. $M_2$Te$_3$ ($M$= Bi, Sb)~\cite{Zhang,Zhu}. A detailed microscopic study of the superconducting state of $\beta$-PdBi$_2$ is essential to investigate the existence of the proposed second gap and more importantly, to detect the influence of any topological aspect in the superconducting pairing symmetry, induced by the topologically protected surface states.


Single crystal samples of $\beta$-PdBi$_2$ were grown from a melt as discussed in Ref.~\onlinecite{Sakano} AC magnetization measurements were performed in a Quantum Design Magnetic Property Measurement System (MPMS). Phase purity of the measured samples was checked using X-ray diffraction measurements. The transverse-field (TF) and zero-field (ZF) $\mu$SR experiments were carried out in the LTF and GPS instruments at the $\pi$M3 beam line of the Paul Scherrer Institute (Villigen, Switzerland). The sample was cooled to the base temperature in zero field for the ZF-$\mu$SR experiments and in 30~mT for the TF-$\mu$SR experiments. Typically $\sim12$~million muon decay events were collected for each spectrum. The ZF- and TF-$\mu$SR data were analyzed using the software package MUSRFIT~\cite{Suter}.


\begin{figure}[htb]
\includegraphics[width=1.0\linewidth]{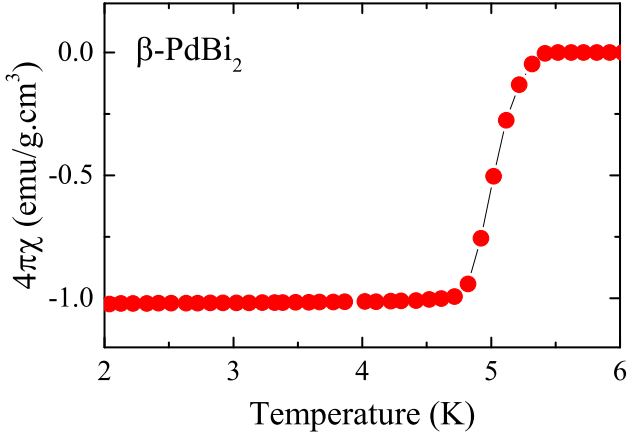}
\caption{(Color online) Temperature dependence of the AC magnetic susceptibility of $\beta$-PdBi$_2$ (real part).}
 \label{fig:magnetization}
\end{figure}

\begin{figure}[htb]
\includegraphics[width=1.0\linewidth]{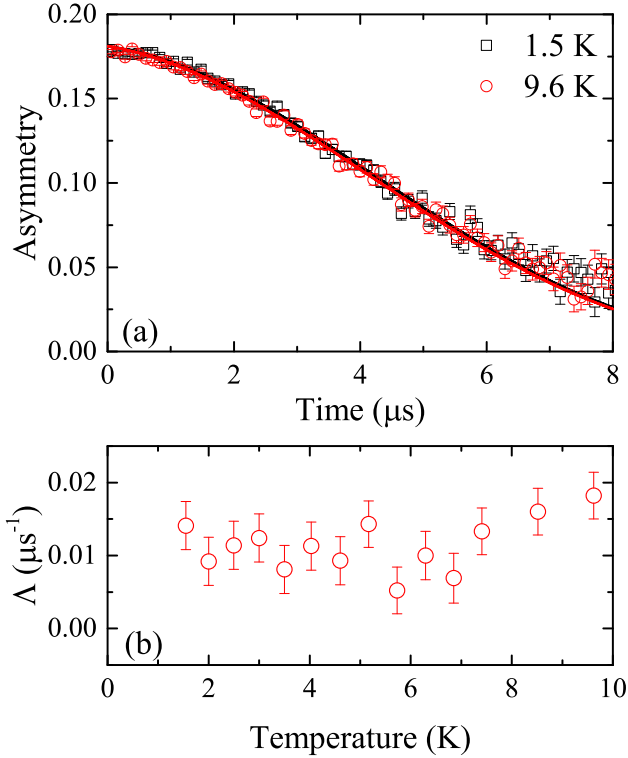}
\caption{(Color online) (a) ZF-$\mu$SR spectra of $\beta$-PdBi$_2$, collected at 9.6 K (above $T_{\rm c}$) and 1.5 K (below $T_{\rm c}$). The solid lines are fits to the data using Eq.~\ref{eq:KT_ZFequation}. (b) shows the temperature variation of the muon spin relaxation rate $\Lambda$, extracted from the ZF-$\mu$SR spectra.}
 \label{fig:AsyZF}
\end{figure}

Figure~\ref{fig:magnetization} shows the real part of the AC magnetic susceptibility of $\beta$-PdBi$_2$. Susceptibility data were corrected by the geometric demagnetization factor. The data display a sharp diamagnetic shift below $T_{\rm c}$ of 5.4 K. The zero-field cooled ac susceptibility approaches a value of −1, showing the sample becomes fully superconducting below $T_{\rm c}$. ZF-$\mu$SR measurements have been performed to study the microscopic magnetic properties in both superconducting and normal states of $\beta$-PdBi$_2$. Figure~\ref{fig:AsyZF}(a) shows the ZF-$\mu$SR signals collected at 9.6 K and 1.5 K. The time dependent ZF-$\mu$SR asymmetry signals above and below $T_{\rm c}$ are consistent, implying that no additional relaxation of the $\mu$SR signal and hence no magnetic moments appear in the superconducting state of $\beta$-PdBi$_2$. However, to confirm that there are no spontaneous internal magnetic fields appearing at $T_c$ associated with e.g., time-reversal-symmetry (TRS) breaking pairing state in $\beta$-PdBi$_2$, ZF-$\mu$SR measurements were carried out over a range of temperature while the sample was heated across $T_{\rm c}$. ZF-$\mu$SR data were analyzed using a combination of Gaussian and Lorentzian Kubo-Toyabe relaxation functions~\cite{Kubo},
\begin{multline}
A(t)= A(0)\left\{\frac{1}{3}+\frac{2}{3}\left(1-a^2t^2-\Lambda{t}\right){\rm exp}\left(-\frac{a^2t^2}{2}-\Lambda{t}\right)\right\}+A_{\rm bg},
\label{eq:KT_ZFequation}
\end{multline}
where $A(0)$ is the initial asymmetry of the sample signal, $A_{\rm bg}$ is the background signal, $a$ and $\Lambda$ are the muon spin relaxation rates due to randomly oriented nuclear moments and diluted electronic moments, respectively. The fits to the ZF-$\mu$SR signals using Eq.~\ref{eq:KT_ZFequation} yield $\Lambda(1.5{\rm K})=0.014(4)$~$\mu$s$^{-1}$, and $\Lambda(9.6{\rm K})=0.018(3)$~$\mu$s$^{-1}$ with a globally free parameter $a=0.191(3)$~$\mu$s$^{-1}$. Figure~\ref{fig:AsyZF}(b) shows the temperature variation of the muon spin relaxation rate $\Lambda$, extracted from the ZF-$\mu$SR spectra. We see essentially no change in relaxation rate over the entire temperature range. This points to the absence of any spontaneous magnetic fields associated with a TRS breaking pairing state in $\beta$-PdBi$_2$, as detected in other superconducting systems~\cite{Luke1,Luke2,Reotier,Aoki,Maisuradze,Hillier,Hillier1,Singh,Biswas}.

\begin{figure}[htb]
\includegraphics[width=1.0\linewidth]{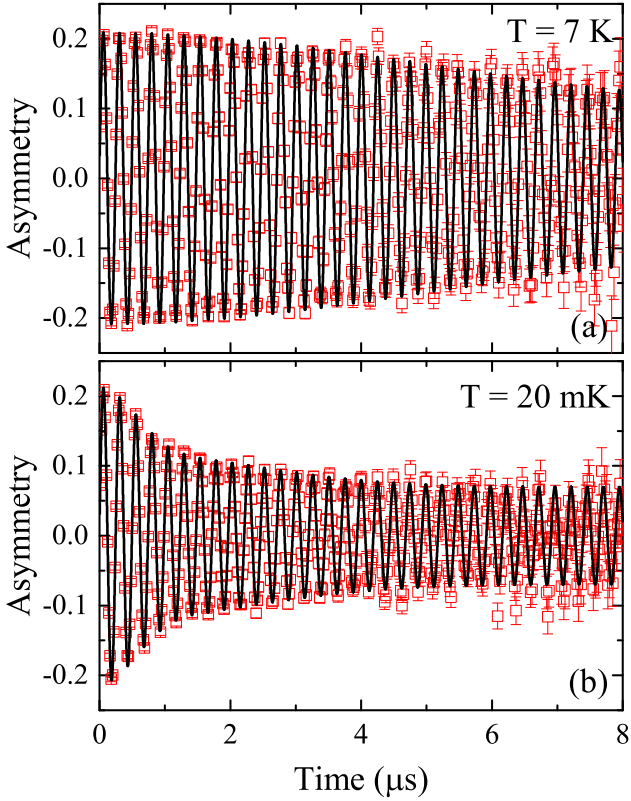}
\caption{(Color online) (a) and (b) show the TF-$\mu$SR time spectra, collected both above and below $T_{\rm c}$ in an applied field of 30 mT. The solid lines are least squares fits according to Eq.~\ref{eq:TFequation}.}
 \label{fig:AsyTF}
\end{figure}

TF-$\mu$SR measurements were carried out to derive the microscopic properties of the superconducting state in $\beta$-PdBi$_2$. Figure~\ref{fig:AsyTF}(a) and (b) display the TF-$\mu$SR time spectra, collected at 7~K and 20~mK under a magnetic field of 30~mT. The $\mu$SR time spectra obtained in the normal state show a much weaker relaxation compared to the one in the superconducting state. This is because, the local field probed by the muons in normal state is produced essentially by the applied field and the signal is only slightly damped due to the presence of nuclear moments in the sample. A more pronounced damping is observed in the superconducting state, caused by the inhomogeneous field distribution generated by the formation of a vortex lattice. The TF-$\mu$SR time spectra were analyzed using the following oscillatory decaying Gaussian function:
\begin{multline}
A^{TF}(t)=A(0)\exp\left(-\sigma^{2}t^{2}\right/2)\cos\left(\gamma_\mu \left\langle B\right\rangle t +\phi\right)\exp\left(-\alpha t\right) \\
+A_{\rm bg}(0)\cos\left(\gamma_\mu B_{\rm bg}t +\phi\right),
\label{eq:TFequation}
\end{multline}
where $A(0)$ and $A_{\rm bg}$(0) are the initial asymmetries of the sample and background signals, $\gamma_{\mu}/2\pi=13.55$~kHz/G is the muon gyromagnetic ratio~\cite{Sonier}, $\left\langle B\right\rangle$ and $B_{\rm bg}$ are the internal and background magnetic fields, $\phi$ is the initial phase of the muon precession signal, $\sigma$ is the Gaussian muon spin relaxation rate, and $\alpha$ is the exponential relaxation rate due to weak and diluted nuclear or electronic moments. $A_{\rm bg}(0)\cos\left(\gamma_\mu B_{\rm bg}t +\phi\right)$ is the background signal that dominantly originates from muons hitting the silver sample holder and was assumed to be non-relaxing over the muon time window. A global fit to the TF-$\mu$SR signals using Eq.~\ref{eq:TFequation} yields $\alpha=0.021(3)$~$\mu$s$^{-1}$. 

\begin{figure}[htb]
\includegraphics[width=1.0\linewidth]{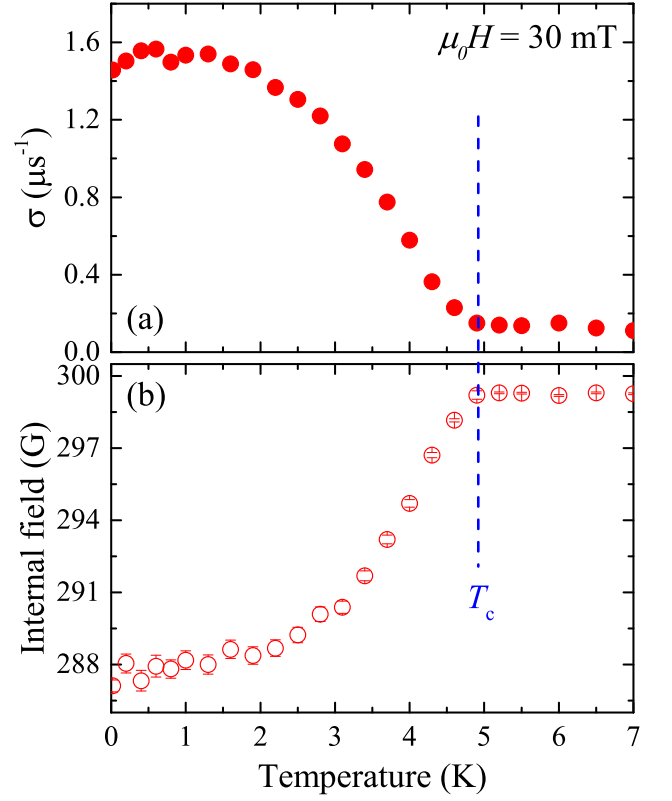}
\caption{(Color online) (a) Temperature dependence of the muon depolarization rate $\sigma$ of $\beta$-PdBi$_2$ collected in an applied magnetic field of 30 mT. (b) Typical diamagnetic shift of the internal field experienced by the muons below $T_{\rm c}$. The dashed line shows the $T_{\rm c}$ value.}
 \label{fig:sigmaT}
\end{figure}

Figure~\ref{fig:sigmaT} (a) reveals the temperature dependence of $\sigma$ of $\beta$-PdBi$_2$ for an applied field of 30 mT, which exhibits a pronounced change at $T = T_{\rm c}$. The temperature dependence of the internal magnetic field at the muon site with the expected diamagnetic shift below $T_{\rm c}$ is shown in Fig.~\ref{fig:sigmaT} (b).

\begin{figure}[htb]
\includegraphics[width=1.0\linewidth]{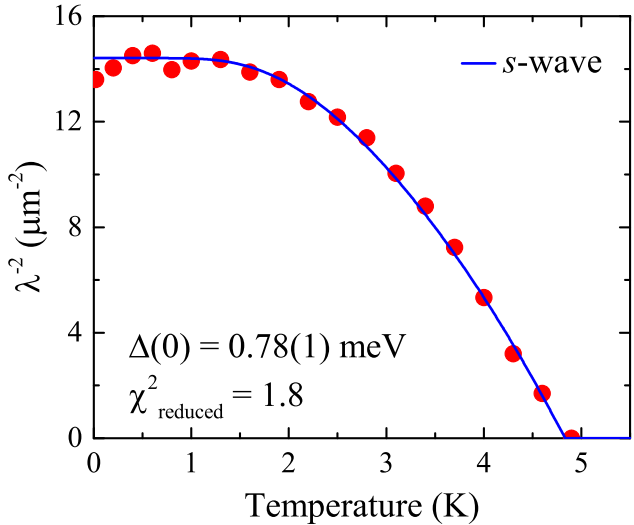}
\caption{(Color online) The temperature dependence of $\lambda^{-2}(T)$. The solid line is a fit to the data with the weak-coupling BCS model.}
 \label{fig:lambdaT}
\end{figure}

The superconducting contribution $\sigma_{sc}$ to $\sigma$ can be derived by quadratically subtracting the nuclear moment contribution $\sigma_{\rm nm}$ (measured above $T_{\rm c}$) as ${\sigma_{\rm sc}}^2=\sigma^2-{\sigma_{\rm nm}}^2$. $\sigma_{\rm nm}$ is assumed to be constant in the temperature range considered here.  In an isotropic type-II superconductor with a hexagonal Abrikosov vortex lattice described by Ginzburg-Landau theory, the magnetic penetration depth $\lambda$ is related to $\sigma_{sc}$ by the Brandt equation~\cite{Brandt}:
\begin{equation}
\sigma_{sc}[\mu{\rm s}^{-1}]=4.854\times10^4\left(1-\frac{H}{H_{\rm c2}}\right)\left[1+1.21\left(1-\sqrt{\frac{H}{H_{\rm c2}}}\right)^3\right]\lambda^{-2}[{\rm nm}^{-2}],
 \label{eq:Brandt_equation}
\end{equation}

where $H$ and $H_{\rm c2}$ are the applied and upper critical field, respectively. The temperature dependence of $\lambda^{-2}(T)$ was calculated using Eq.~\ref{eq:Brandt_equation} where $H_{\rm c2}(T)$ was obtained from Ref.~\onlinecite{Imai} by averaging over the values collected parallel to \textit{ab}- and \textit{c}-axis. Figure~\ref{fig:lambdaT} shows $\lambda^{-2}(T)$ which is proportional to the effective superfluid density, $\rho_s\propto\lambda^{-2}$. This provides a possibility to study the gap symmetry in the electronic density of states in the proximity of the Fermi energy below $T_{\rm c}$  in $\beta$-PdBi$_2$. Figure~\ref{fig:lambdaT} reveals that the $\rho_{\rm s}$ is nearly constant below $T_{\rm c}/3\approx2$ K, suggesting a nodeless superconducting gap in $\beta$-PdBi$_2$. A reasonable fit to $\lambda^{-2}(T)$ can be achieved with a single-gap BCS $s$-wave model (solid blue line in Fig.~\ref{fig:lambdaT})~\cite{Tinkham,Prozorov}:

\begin{equation}
\frac{\lambda^{-2}(T)}{\lambda^{-2}(0)}= 1+2\int_{\Delta(T)}^{\infty}\left(\frac{\partial f}{\partial E}\right)\frac{E dE}{\sqrt{E^2-\Delta(T)^2}}.
 \label{eq:lambda}
\end{equation}

Here $\lambda^{-2}(0)$ is the  zero-temperature value of the magnetic penetration depth and $f=[1+\exp(E/k_BT)]^{-1}$ denotes the Fermi function. The BCS temperature dependence of the superconducting gap function is approximated as~\cite{Carrington}
\begin{equation}
\Delta(T)=\Delta(0)\tanh\left\{1.82\left[1.018\left(\frac{T_{\rm c}}{T}-1\right)\right]^{0.51}\right\},
 \label{eq:delta}
\end{equation}

where $\Delta(0)$ is the gap magnitude at zero temperature. From the fit, we obtain $T_{\rm c}=4.84(1)$~K, $\lambda(0)=263(10)$~nm, and $\Delta(0)=0.78(1)$~meV. The gap to $T_{\rm c}$ ratio $\Delta(0)/k_{\rm B}T_{\rm c}=1.87(3)$ is slightly higher than the BCS value of 1.76. Our results are consistent with the reported values on $\beta$-PdBi$_2$, measured using STM by Herrera \textit{et al.} \cite{Herrera}.

\begin{figure}[htb]
\includegraphics[width=1.0\linewidth]{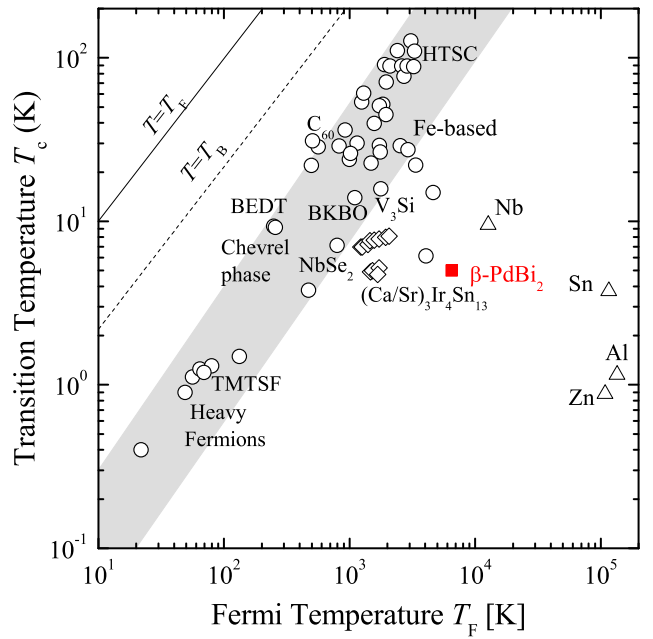}
\caption{(Color online) The Uemura plot, showing the superconducting transition temperature $T_{\rm c}$ vs the effective Fermi temperature $T_{\rm F}$ for $\beta$-PdBi$_2$ is shown as solid red square. Other data points plotted in the $T_{\rm c}$ vs $T_{\rm F}$ diagram represent the different families of superconductors (plot adapted from \onlinecite{Biswas1,Hashimoto,Khasanov}). The dashed line corresponds to the Bose-Einstein condensation temperature $T_{\rm B}$.}
 \label{fig:UemuraPlot}
\end{figure}

To put our finding on $\beta$-PdBi$_2$ in the context of other superconductors, we plot the Uemura graph of $T_{\rm c}$ versus effective Fermi temperature $T_{\rm F}$ (see Fig.~\ref{fig:UemuraPlot}) which is often used to define the character of unconventionality of a superconductor \cite{Hashimoto,Uemura}. $T_{\rm F}$ for $\beta$-PdBi$_2$ was calculated using the 3D expression as discussed in Ref.~\onlinecite{Biswas1}. We estimate $T_{\rm F}=6497$ K for  $\beta$-PdBi$_2$, which places this material well outside the broad gray line for unconventional superconductors, such as cuprates and iron-based superconductors. Conventional elemental superconductors are also generally considered to lie on the right-hand side of this diagram. This provides further evidence for conventional BCS $s$-wave superconductivity in $\beta$-PdBi$_2$.

Using the values of $\lambda$ and $\xi$ (taking the average value of $\xi_{\rm c}$ and $\xi_{\rm ab}$ from \onlinecite{Kacmarcik}), we calculate the GL parameter, $\kappa=\frac{\lambda}{\xi} = 12.6(5)$. By combining the value of $\xi$ and our measured value of $\lambda$, we calculate the lower critical field, $H_{\rm c1}$ using the expression~\cite{Brandt}:
\begin{equation}
\mu_{0}H_{\rm c1}=\frac{\phi_0}{4\rm \pi\lambda^{2}}\left(\ln \frac{\lambda}{\xi}+0.5\right)
 \label{eq:hc1}
\end{equation}
and obtain $\mu_{0}H_{\rm c1}(0)=6.1(3)$~mT.


In conclusion, $\mu$SR studies have been performed on the superconducting $\beta$-PdBi$_2$. ZF-$\mu$SR data show no sign of a magnetic anomaly in the superconducting ground state of $\beta$-PdBi$_2$. TF-$\mu$SR measurements were performed to study the temperature dependence of the London penetration depth $\lambda$ which is proportional to the superfluid density $\rho$. $\lambda (T)$ is very well modeled within a single-gap BCS $s$-wave scenario with $\Delta(0)=0.78(1)$~meV, provide well-founded evidence for nodeless superconductivity in $\beta$-PdBi$_2$. The magnetic penetration depth was estimated as $\lambda(0)=263(10)$~nm. The value of $\lambda^{-2}(0)$ places $\beta$-PdBi$_2$ well outside the broad line for unconventional superconductors in a Uemura plot. Our results further suggest that the topologically protected surface states have very little or no effect on the bulk of the superconductor. It is important to mention here that in the bulk-$\mu$SR technique muons penetrate typically a few hundreds of $\mu$m with a range straggling of the order of $\sim$ 100 $\mu$m. By contrast, we expect the spatial extension of the surface Andreev bound states to be of the order of the coherence length, i.e. a few tens of nm. Further studies using surface probes are therefore required to understand the evolution of the superconductivity at the surface. By the time we submitted our manuscript, we noticed a new article on this material in arXiv.org by J. Ka\v{c}mar\v{c}\'{i}k \textit{et al.}~\cite{Kacmarcik} in which they report calorimetric, Hall-probe magnetometry and STM measurements. They also found standard single \textit{s}-wave gap superconductivity, consistent with our finding.

The $\mu$SR experiments were performed at the Swiss Muon Source (S$\mu$S), Paul Scherrer Institute (PSI, Switzerland).

\end{document}